# Extending the analogy between intracellular motion in mammalian cells and glassy dynamics

B. Corci,[1,2] O. Hooiveld,[1] A. M. Dolga,[1] C. Åberg[1]

[1] Groningen Research Institute of Pharmacy, University of Groningen, Antonius Deusinglaan 1, 9713 AV Groningen, The Netherlands.
[2] Zernike Institute for Advanced Materials, University of Groningen, Nijenborgh 4, 9747 AG Groningen, The Netherlands.

**Abstract**
The physics of how molecules, organelles, and foreign objects move within living cells has been extensively studied in organisms ranging from bacteria to human cells. In mammalian cells, in particular, cellular vesicles move across the cell using motor proteins that carry the vesicle down the cytoskeleton to their destination. We have recently noted several similarities between the motion of such vesicles and that in disordered, "glassy", systems, but it remains unclear whether that is a general observation or something specific to certain vesicles in one particular cell type. Here we follow the motion of mitochondria, the organelles responsible for cell energy production, in several mammalian cell types over timescales ranging from 50 ms up to 70 s. Qualitative observations show that single mitochondria remain stalled, remaining within a spatially limited region, for extended periods of time, before moving longer distances relatively quickly. Analysing this motion quantitatively, we observe a displacement distribution that is roughly Gaussian for shorter distances (≲ 0.05 µm) but which exhibits exponentially decaying tails at longer distances (up to 0.40 µm). We show that this behaviour is well-described by a model originally developed to describe the motion in glassy systems. These observations are extended to in total 3 different objects (mitochondria, lysosomes and nano-sized beads enclosed in vesicles), 3 different mammalian cell types, from 2 different organisms (human and mouse). We provide further evidence that supports glass-like characteristics of the motion by showing a difference between the time it takes to move a longer distance for the first time and subsequent times, as well as a weak ergodicity breaking of the motion. Overall, we demonstrate the ubiquity of glass-like motion in mammalian cells, providing a different perspective on intracellular motion.

**INTRODUCTION**
The living cell is constantly active with a myriad of processes simultaneously occurring at the same time within it. Some of these processes depend intimately upon transport, whether of molecules, molecular complexes, or larger lipid bilayer-decorated vesicles. While it appears that many molecular species are transported sufficiently rapid by diffusion[1], for many vesicles (and the cargo they enclose) the cell has evolved a highly regulated transport system composed of a polymer network (the cytoskeleton) upon which the vesicles are transported, being carried by motor proteins that consume cell energy each step they take[2].

We recently followed the motion of such intracellular vesicles in human (HeLa) cells using fluorescent nano-sized beads as labels[3]. These beads populate vesicles[4–6] (many of them lysosomes)[4–8] that can be transported by motor proteins. We observed that the motion of the vesicles (beads) exhibited several hallmarks that are characteristic of the motion in disordered, "glassy", systems[9]. As opposed to diffusive motion[10], the motion in glassy systems is characterised by a highly heterogeneous dynamics where, roughly speaking, the majority of objects spend most of their time stationary and when they do move, they do so in a highly collective manner[11]. Glass-like motion was later reported also for chloroplasts in plant cells[12]. A similarity between the motion in glassy systems had already been noted for the motion in bacteria[13] and other microorganisms[14], but the physical interpretation is likely quite different, since bacteria do not have motor proteins. Indeed, motor protein-driven transport is generally understood to exist to *speed up* transport[2], so the observation that it shares characteristics with the notoriously sluggish motion in a glass may be considered somewhat surprising.

To test the generality of the observation that intracellular motion is glass-like we here followed the motion of mitochondria in three different cell types from two different organisms (human and mouse). Mitochondria are double-membrane organelles which are essential for all eukaryotic cells. While their main function is to supply cell-energy, they are involved in multiple intracellular processes such as calcium storage, reactive oxygen species formation, and apoptosis. Mitochondrial shape, as well as the distribution of mitochondria within the cell, is highly heterogeneous due to the dynamic nature of these organelles. Most mitochondria are present as tubular elongated structures fused together to form a widespread network, which co-exists with single mitochondria detached from the network[15]. The simultaneous presence of different mitochondrial configurations in the cell is due to the constant remodelling of the mitochondrial network through fission and fusion processes, mainly driven by the metabolic needs of the cell and interactions with other organelles and cytoskeleton structures[16]. Crucially for our purposes here, mitochondria are transported around inside the cell along the microtubule network, being carried towards the plus ends (typically towards the cell periphery) by kinesin motor proteins, and towards the minus ends (typically towards the cell body) by dynein motor proteins[16–18]. We thus use them as another example of organelles that move by motor proteins and study their motion using live-cell fluorescence microscopy. Analysing their trajectories we test several hallmarks characteristic of the motion in glassy systems and show that, indeed, mitochondria also move in a glass-like manner.

**MATERIALS AND METHODS**

**Cell culture**
HEK 293 (American type culture collection, ATCC; no. CRL-1573, lot no. 63966486)) cells were cultured in Dulbecco's Modified Eagle's Medium (DMEM; Gibco, Life Technologies, Eugene, OR, USA), supplemented with 10% foetal bovine serum (FBS; Gibco, Life Technologies, Eugene, OR, USA). The cells were kept in a 37°C and 5% $CO_2$ incubator and were subcultured twice per week. Cells were used at passage number 6–26. HeLa cells (ATCC; no. CCL-2TM, lot no. 61647128) were cultured under the same conditions and subcultured thrice per week. Cells were used at passage number 16–30. HT22 cells were cultured in DMEM (Gibco, ThermoFisher

Scientific, Landsmeer, The Netherlands) with 10% FBS (GE Healthcare Life Sciences, Eindhoven, the Netherlands), 1% sodium-pyruvate (ThermoFisher Scientific, Landsmeer, The Netherlands) and 2% penicillin-streptomycin (ThermoFisher Scientific, Landsmeer, The Netherlands), kept in a 37°C and 5% $CO_2$ incubator, and subcultured twice per week. Cells were used at passage number 270. Regular mycoplasma tests were carried out and only mycoplasma negative cells were used for the experiments. For the microscopy experiments, 150,000 (HEK 293) and 50,000 (HeLa and HT22) cells were seeded onto petri dishes with glass-bottom microwells (MatTek, Ashland, MA, USA).

**Organelle staining and live cell imaging microscopy**
The experiments were conducted 24 h after cell seeding. To fluorescently label the mitochondria, cells were treated with MitoTracker Deep Red (Invitrogen, Waltham, MA, USA) diluted in serum-free DMEM to a concentration of 200 nM and incubated for 25 min (37°C, 5% $CO_2$). To fluorescently label the lysosomes, cells were treated with Lysotracker Red (Invitrogen, Waltham, MA, USA) diluted in DMEM to a concentration of 0.75 µM and incubated for 1 h (37°C, 5% $CO_2$). After incubation with the relevant stain, the cells were washed three times with phosphate-buffered saline solution (Gibco, Life Technologies, Eugene, OR, USA), after which transparent live cell imaging solution (Invitrogen, Waltham, MA, USA) was added. For the experiment on fixed cells, the cells were treated with paraformaldehyde (4%) for 15 min at room temperature before microscopy. In general, microscopy was performed directly after staining. Cells were observed with a DeltaVision Elite inverted microscope (based on an Olympus IX-71 stand) using a 60× oil objective, illuminating the cells with a solid state illumination system. For MitoTracker a 621–643 nm excitation and a 662–696 nm emission filter was used, while for LysoTracker a 528–555 nm excitation and a 574–619 emission filter was used. Images were captured with a PCO-edge sCMOS camera every 50 ms for 70 s. The same settings were used for all three cell types.

**Organelle tracking**
To track single mitochondria, a region of interest within a cell was selected to isolate the mitochondrion. The trajectories were then automatically identified using the TrackMate plugin (version 2.8.1), available in the Fiji distribution of ImageJ. After the automatic tracking all trajectories were checked manually. The same procedure was used for lysosome tracking.

**Analysis**
The analysis performed here follows that in our previous work[3]. Briefly, the mean square displacement was determined by calculating the square displacement (in two dimensions) between all pairs of time points along the length of the trajectory, and subsequently averaging this over all pairs with the same lag time and over all trajectories, thereby forming a time- and ensemble-averaged mean square displacement.

The displacement distribution was calculated starting from the radial displacement, Δr, again between all pairs of time points along the length of all trajectories. Subsequently, the histogram of all radial displacements for a given time duration was determined and finally normalised such that the integral of the distribution multiplied by 2πΔr over Δr was 1. We fitted the Chaudhuri *et al.* model[11,19] to the experimentally determined displacement distributions, using the two-dimensional version of the model as written down in our previous work (Eq. S8 of ref. 3). This model depends on four parameters ($\tau_1$, $\tau_2$, $l$ and $d$; see main text for their definition) and the fit was performed on the displacement distribution for several different lag times *simultaneously* (a global fit). The resulting fitting parameters are shown below (Table 1 of the Result section).

The time that a particle moves for the first time a certain distance and the time that it takes to move the same distance a second time were calculated as described in previous literature[20]. In doing so, we considered all pairs of time points, to improve statistics, and limited the calculation to the first 10 s, to reduce bias.

The degree of ergodicity breaking as a function of lag time, τ, was quantified through the ergodicity breaking parameter[21–23]

$$\mathrm{EB}(\tau) = \lim_{t \to \infty} \frac{\langle X^2 \rangle - \langle X \rangle^2}{\langle X \rangle^2} \equiv \lim_{t \to \infty} \mathrm{EB}_t(\tau)$$

where $X$ stands for the mean square displacement of an individual trajectory at a given lag time, τ, averaged along the trajectory up to time $t$, and the brackets denote ensemble averaging. Since it is not possible to measure the infinite time limit experimentally, we evaluated the time-dependent parameter $\mathrm{EB}_t(\tau)$ defined above and observed how it changed with time.

**RESULTS AND DISCUSSION**

We used Human Embryonic Kidney (HEK 293) cells as our main cell model, with selected experiments repeated using HeLa (adenocarcinomic human cervical epithelial) cells and HT22 (mouse hippocampal neuronal) cells. To visualise and characterise mitochondrial motion, the mitochondria were stained with Mitotracker Deep Red and subsequently observed using live-cell fluorescence microscopy. We performed the microscopy in widefield mode, which implies a two-dimensional view of the cell, but much faster acquisition times. With a fast-acquisition camera, we were able to acquire images of the cells every 50 ms, for a total of 1400 frames or 70 s.

Mitochondria are dynamic organelles present throughout the cell in different dimensions and shapes. Their diameter can vary from 0.5 to 1 μm and they can exhibit a more or less elongated tubular structure that can extend several μm in length[24,25]. The detailed mitochondrial structure is difficult to visualise through optical microscopy, due to their dimension being close to the diffraction limit of visible light. Nevertheless, it is possible to get a general overview of the structure and this is sufficient for our purposes here. Thus, we observe that the majority of mitochondria appear connected in a network, while some mitochondria instead appear as punctate "spots" (Figure 1, right panel). These single (punctate) mitochondria are mostly present in the nuclear region (presumably under the cell nucleus, since mitochondria do not enter the nucleus) and closer to the outer cell membrane. In contrast, the mitochondrial network is spread out in the central part of the cell, surrounding the cell nucleus.

We chose to focus on the motion of the "single" (punctate) mitochondria as being more representative of general organelle motion than the mitochondrial network (however, see previous literature for approaches to characterise the network[26,27]). To characterise the motion of these single mitochondria, we determined trajectories from the microscopy images by automatic tracking using the open-source software TrackMate[28], a Fiji/ImageJ plugin[29–30], reviewing the accuracy of each trajectory manually afterwards. Some mitochondria (*e.g.*, Figure 1 left panels) could be followed throughout the entire duration of the "movie" (1400 frames, 70 s). However, not all of the mitochondria were easily detectable for such a long time, either because they moved to a different focal plane (not observable in two-dimensional microscopy) or because they were obscured by the widespread mitochondrial network. We thus followed the mitochondria for as long as they were visible, resulting in a distribution of trajectory lengths (Supplementary Figure 1).

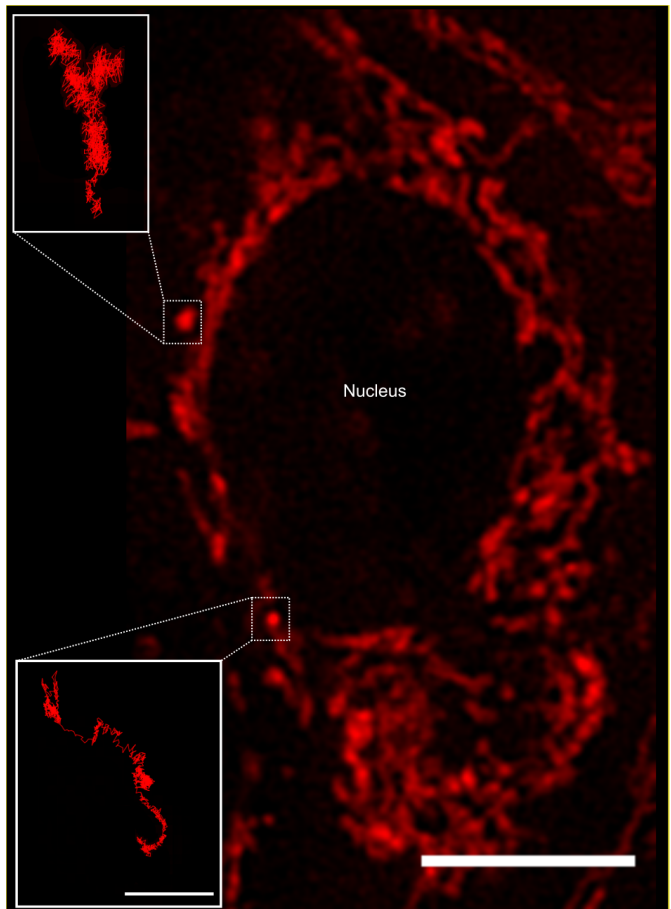

**Figure 1: Mitochondria trajectories.** (Right) Fluorescence microscopy image of a HEK 293 cell stained with MitoTracker deep red for visualising the mitochondria. The central region is occupied by the nucleus (not stained and hence only indirectly visible due to the absence of mitochondria), around which a mitochondrial network is present. Additionally, some mitochondria appear more punctate in nature and isolated from the network, as indicated by the two examples. Here we focussed on the motion of these single mitochondria. Scale bar 10 μm. (Left) Trajectories of the two single punctate mitochondria, tracked as described in Materials and Methods. One observes phases of (nearly) unidirectional motion, interspersed by moments when the mitochondrion does not move very far. Scale bar 1 μm.

Tracked mitochondria show trajectories of sometimes several micrometres in length (Figure 1, left panels). Their motion appears highly heterogeneous, exhibiting moments in which the mitochondrion is "rattling" around in the same general area, interspersed by longer "jumps" with a considerable degree of directionality. The time spent rattling around in the same area is often longer (in both of the trajectories shown in Figure 1 it is around 22 s) while the duration of the jumps are typically faster (around 8 s in the Figure 1 trajectories). The unidirectionally of the jumps most likely reflect active motor protein-driven motion. Nevertheless, we note that the qualitative appearance of the trajectories show a striking similarity with the motion in a number of glassy systems, such as a colloidal glass[31], a sheared granular material[32], and a glass-forming binary Lennard-Jones mixture[11]. It is this similarity between intracellular motion and glassy motion[3] that we want to explore further here.

As a first characterisation of mitochondrial motion, we evaluated the mean square displacement (Figure 2). To improve statistics, the calculations were done averaging all collected trajectories over all pairs of time points, *i.e.*, we calculated the both time- and ensemble-averaged mean square displacement. To check the accuracy of our measurements, we started by performing the same type of experiment as in live cells, but on fixed cells (cells that are no longer alive, but with preserved intracellular structure). In these cells, the mitochondria cannot move by active processes, allowing us to assess eventual background movement due to drift of the microscope stage and thermal diffusion, as well as the finite precision with which we can localise a mitochondrion. These control experiments show (red curve) almost no displacement, with only a small increase at longer timescales of around 20–30 s. Furthermore, it may be observed that for shorter timescales the curve plateaus at an $\langle r^2 \rangle$ value of ~0.0012 µm$^2$ or a distance of ~$\sqrt{0.0012}$ µm$^2$ ≈ 0.035 µm. Whether to interpret this as localisation imprecision[33,34] or genuine short scale motion is less clear. Importantly, though, this distance is shorter than the typical length scales of interest to us here.

Turning to our real object of interest, the mitochondria in live cells, the mean square displacement (Figure 2, black curve) shows an initial plateau at an $\langle r^2 \rangle$ value of ~0.0031 µm$^2$ or a distance of ~$\sqrt{0.0031}$ µm$^2$ ≈ 0.056 µm. The plateau is followed by a slow increase with time for short time scales (below 1 s), after which it grows as a power law with exponent 1.15. After ~10 s there is possibly a decrease in the power law exponent, but it should be noted that the statistics is poorer here. The initial plateau is qualitatively consistent with the observations made above (Figure 1, left panels) and would then represent the rattling around of the mitochondrion in the same general area, with an average extent of 0.056 µm. Subsequently, at a timescale of ~200 ms some mitochondria start performing longer jumps and the mean square displacement starts increasing. As more and more mitochondria perform jumps and/or the mitochondria that have already jumped jump again, the mean square displacement increases further.

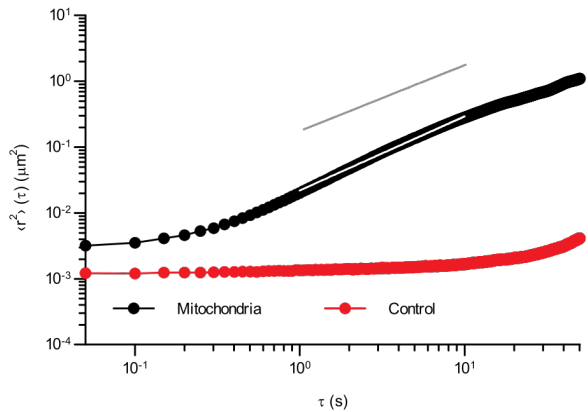

**Figure 2: Mean square displacement of mitochondria in HEK 293 cells.** The mean square displacement was calculated from the trajectories, averaging both over time and trajectories (ensemble). Error bars represent standard error of the mean, but are too small to be visible. The data was limited to the first 50 s (of 70 s) as the statistics get increasingly poorer for longer times. (Black) Mitochondrial motion in live cells (84 trajectories). (Red) Mitochondrial motion in fixed cells as a control (20 trajectories). (White solid line) A power law fit to the live-cell data in the interval 1 to 10 s, resulting in an exponent of 1.15. (Grey solid line) Slope of 1, representative of Brownian motion, as a comparison. *N. B.* the log-log scale.

To better characterise the heterogeneity of the motion, we next calculated the displacement distribution. The displacement distribution (also referred to as the self-part of the van Hove function[35,36]) describes the probability that a particle moves a distance Δr in time $\tau$ from its starting point at time 0. Like for the mean square displacement, we evaluated it for all trajectories and for all pairs of time-points, and present it for a few selected lag times, $\tau$, between 0.5 and 3.0 s (Figure 3; see also Supplementary Figure S2 for more times). A first observation is that the displacement distribution is clearly not Gaussian, but rather exhibits exponentially decaying tails (Figure 3; note the *y* axis log scale). This is most vividly demonstrated by explicitly fitting a Gaussian distribution to the data which, indeed, completely fails to describe the tails (Supplementary Figure S3). Considering that mitochondria are transported by motor proteins along the cytoskeleton, it is perhaps not surprising that the motion is non-Gaussian. Indeed, in general multiple examples in the literature have reported a non-Gaussian behaviour of intracellular motion[37–39]. Still, it is telling that even at the longer timescales of a few seconds, when the mean square displacement increases close to linearly (Figure 2, white line), the displacement distribution remains non-Gaussian.

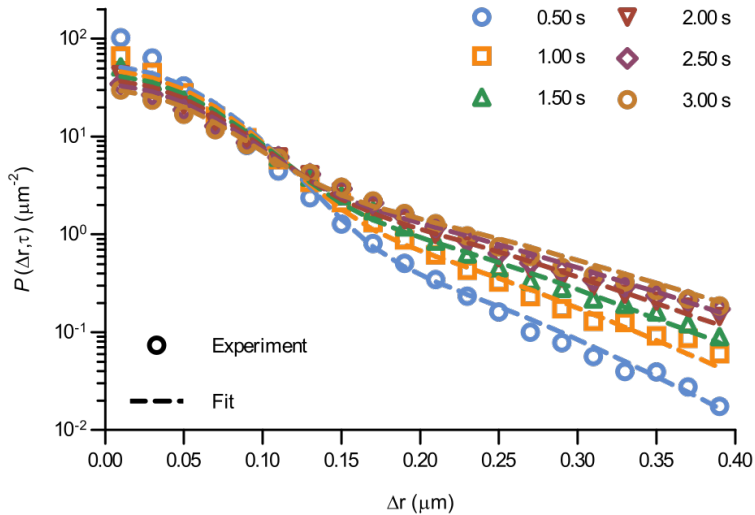

**Figure 3: Displacement distribution of mitochondria in HEK 293 cells.** (Data points) Displacement distributions of mitochondria observed experimentally for different lag times. (Dotted lines) Fits of a model describing glassy motion[11,19] to the experimental data. The four fitting parameters ($\tau_1$, $\tau_2$, $l$, $d$) were ensured to be the same for all lag times and are reported in Table 1. For a better visualisation of the data only selected times are shown here, but more lag times were included in the fit (see Supplementary Figure S2 for a complete view).

Even more telling is the fact that these displacement distributions (Figure 3) are rather similar to those observed in a range of glassy systems[11,19,31,32,40–42], as we have previously noted for beads being moved by vesicles in HeLa cells[3]. Indeed, this is consistent with the similarity between trajectories such as those shown in Figure 1 and those reported for "real" glassy systems[11,31,32]. To evaluate if this similarity is not just qualitative but *quantitative*, we used the model introduced by Chaudhuri *et al*[11,19]. Their model describes the displacement distribution in several different experimental[19,31,32] and simulated[40–42] systems exhibiting glassy dynamics, and is based upon the following picture: Particles spend some time rattling around locally, without moving larger distances. This part of the motion is described with a (time independent) Gaussian displacement distribution with characteristic distance, $l$. Every now and then the particle subsequently takes a longer jump, described by another Gaussian with a different characteristic distance, $d$. How long the particle spends rattling around before making a jump is taken from an exponentially decaying waiting time distribution with a characteristic time. To mimic "ageing", the characteristic time before taking the first jump (the persistence time), $\tau_1$, is allowed to be different from the characteristic time before taking a subsequent jump (the exchange time), $\tau_2$. For the glassy systems the model was originally applied to, the reason the particle remains stalled before taking a longer jump is related to the denseness of the system (particles blocking each other from moving). We remain agnostic as to whether this is also the mechanism underlying the organelle motion and use the model only because of the similarity between organelle motion and that observed in glassy systems.

To test this model, we performed a global fit (where the parameters were ensured to be the same for the different time points analysed) to the experimentally observed displacement distributions. We observe that such a fit describes the data well (Figure 3 solid lines; see also Supplementary Figure S2 for more time points) demonstrating that glassy motion is a good model to describe individual mitochondrial motion in these cells. In terms of the characteristic distances extracted from the fit (Table 1), we find that the rattling around occurs at a length scale of $l$ = 0.05 µm, consistent with the plateau of the mean square displacement (Figure 2 black). Similarly, we find a length scale of the jumps of $d$ = 0.10 µm. Both lengths are much smaller than the average size of the (single, punctate) mitochondria.

In terms of the time it takes before performing a jump, we observe that the fit results in a time before taking a first jump, $\tau_1$, that is different from the time before taking a subsequent jump, $\tau_2$ (Table 1). Indeed, it is a typical feature of glassy systems that the distributions of these two times are different (decoupled). To provide further evidence of this observation, we proceed by evaluating the distribution of these two times directly from the experimental data (rather than indirectly from the fits). Thus, we calculated the two time distributions up to 10 s, defining the size of a "jump" as 0.10 µm, since this was the size of a jump extracted from the fit (Table 1). From this calculation it is clear (Figure 4) that the time before making the first jump (persistence time; blue) is different from the time before making a subsequent jump (exchange time; orange). The same observation is made for other distances defining a "jump" ranging from 0.05 to 0.20 µm (Supplementary Figure S4). For mitochondria this means that the probability to stop rattling around and perform a jump is higher after a first jump has already been performed. In a "real" glass, this is related to the probability of empty space becoming available for a particle to move into and the concept of dynamic heterogeneity. For the mitochondria of relevance here it is likely related to the processivity of motor proteins, which implies that once the organelle is moving it is biased to keep moving. Either way, the validation of this hallmark of glassy dynamics reinforces our previous finding of a model of glassy motion providing a good description of mitochondrial motion.

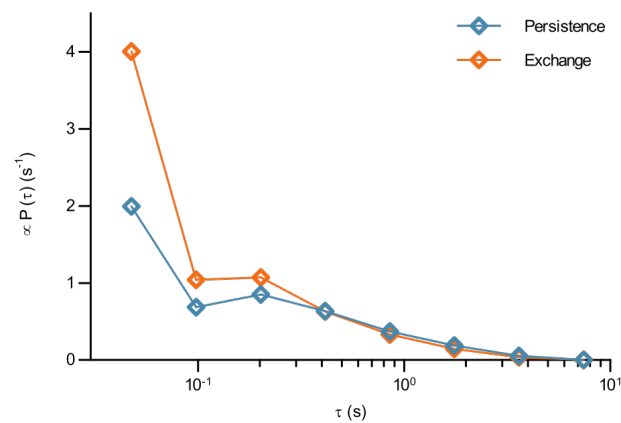

**Figure 4: The waiting time before jumping the first time is different from the waiting time before jumping again for mitochondria in HEK 293 cells.** How long it took a mitochondrion to move a distance of 0.1 µm for the first time, and the time it took to move the same distance a second time having already done so once, was evaluated from the experimental data as described in Methods. The distribution of (Blue) the time it takes to move the first time (persistence time) and (Orange) the time it takes to move a subsequent time (exchange time). The distributions for a few other choices of the jump length are reported in Supplementary Figure S4. Note the log scale.

Up to this point we have demonstrated the utility of describing the motion of mitochondria in HEK 293 cells in terms of concepts borrowed from glassy dynamics. In order to expand the concept of intracellular motion being glassy to other systems, we first considered other cell lines. Thus, mitochondrial motion was assessed, following the same procedure as before, in HeLa cells and HT22 cells. HeLa cells are extensively used in cell biological research, are of epithelial nature and are, like HEK 293 cells, of human origin. HT22 cells, in contrast, are not human but of murine origin, and belong to an immortalised hippocampal neuronal cell line. In addition, we also wanted to see whether the motion of other organelles could be described in the same way, so we also tracked lysosomes in HEK 293 cells. Like mitochondria, lysosomes are membrane-bound organelles in mammalian cells, heavily involved in metabolic processes, in particular in the degradation of cell products[43]. For all of these different systems we calculated the time and ensemble-averaged mean square displacement (Figure 5 top row), as a first evaluation of the type of motion, and also fitted the Chaudhuri *et al.* model to the experimentally determined displacement distributions (Figure 5 bottom row). Finally, as a further comparison, we also reproduce our previously published results[3] on beads being moved by vesicles in HeLa cells (Figure 5d,h).

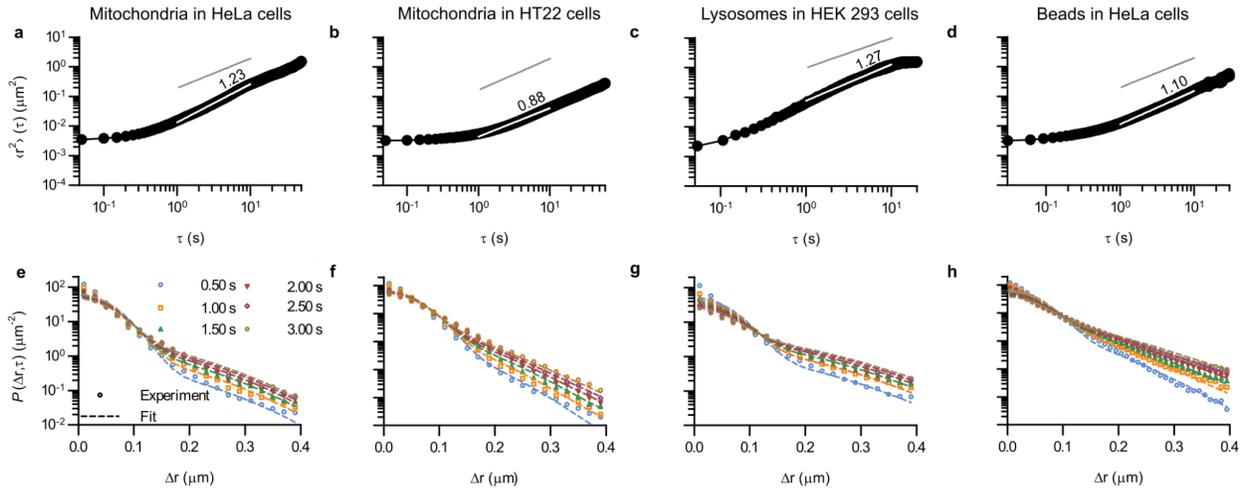

**Figure 5: Ubiquity of glass-like characteristics of organelle motion in mammalian cells.** (a–d) Mean square displacement as a function of lag time, τ. (e–h) Displacement distribution for a few selected lag times. (Dotted lines) Fits of a model describing glassy motion[11,19] to the experimental data. The four fitting parameters ($\tau_1$, $\tau_2$, $l$, $d$) were ensured to be the same for all lag times and are reported in Table 1. (a,e) Mitochondria in HeLa cells; (b,f) Mitochondria in HT22 cells; (c,g) Lysosomes in HEK 293 cells; and (d,h) Polystyrene beads in HeLa cells (data reproduced from literature[3]). Trajectory length distributions are shown in Supplementary Figure S5.

The mean square displacement in all of the different systems exhibit a largely similar behaviour (Figure 5 top row), with an initial plateau for short times, followed by a slow increase until ~ 1 s, after which it grows more rapidly. At longer times, most of the systems have a slightly superdiffusive trend, though for mitochondria in HT22 cells the slope is actually below 1 (Figure 5b). Aside from the qualitative similarities, it is also interesting to note a surprising amount of quantitative overlap. Thus the initial, 50 ms, value of the mean square displacement of the different objects in different cells is around 0.003 µm$^2$ for all tracked organelles in all cell types (Figure 5a–b,d and Figure 2), though for lysosomes it is slightly lower at around 0.002 µm$^2$ (Figure 5c). Contributing to this value is localisation imprecision, which we have already argued is not the main factor based on the value we found for mitochondria in fixed HEK 293 cells (Figure 2, red curve). Different objects (mitochondria, lysosomes and beads), presumably exhibit various ease of localisation due to their different size and fluorescence properties. Furthermore, the present work (Figure 2 and Figure 5a–c) and our previous results reproduced here (Figure 5d) were acquired using different microscopy set-ups, which one again would expect would give rise to a difference in localisation precision. The fact that we nevertheless find similar values under these variations thus gives further evidence that the initial plateau represents (mainly) genuine motion. In terms of quantitative agreement it is even more encouraging to note that if we focus on HeLa cells specifically, then the mean square displacement of mitochondria and beads completely overlap for shorter times (Supplementary Figure S6).

Also for the displacement distribution (Figure 5 bottom row and Figure 2) there is a qualitative agreement between all the different systems. However, we also note agreement of a quantitative nature, namely that the Chaudhuri *et al.* model, originally developed to describe the motion in glassy systems, describes the displacement distribution in all of the different systems well. We have thereby enlarged the scope of this analogy substantially.

In terms of the parameters extracted from the fits (Table 1) we observe no clear picture when it comes to the two waiting times. However, the range of the rattling, $l \approx 0.05$ µm, is remarkably similar for all the systems. This is perhaps somewhat surprising given one might think it would be related to the size of the organelles, which are a bit different. Furthermore, the jump distance, $d \approx 0.10$ µm, is also notably similar for all systems.

While one should perhaps not stretch the analogy too far, we nevertheless find it interesting to note that for all the systems we find that the range of the rattling, $d$, is roughly equal to twice the jump distance, $l$, ($d \approx 2l$; compare last two columns). This is the relation already noted by Chaudhuri *et al.*[11] to be approximately fulfilled for a colloidal glass[31], a sheared granular material[32], a glass-forming binary Lennard-Jones mixture[40], and an attracting colloidal gel[19]. Whether this is a coincidence or whether it hints at a similar mechanistic basis we remain agnostic about.

**Table 1: Parameters extracted from fits to the displacement distributions.** The Chaudhuri *et al.* model[11,19] describing the motion in glassy systems was fitted to the experimental data (Figure 2 and Figure 5 bottom row). The model is formulated in terms of four different parameters: $\tau_1$, the waiting time before the organelle makes its first jump; $\tau_2$, the waiting time before any subsequent jump; $l$, the spatial extent of the organelle rattling around; and $d$, the size of the jump. For each system, (first column in the table) the parameters were ensured to be the same for all different lag times. The last column tests the relationship between $l$ and $d$ parameters observed to be approximately fulfilled by Chaudhuri *et al.*[11]

| System | $\tau_1$ (s) | $\tau_2$ (s) | $l$ (µm) | $d$ (µm) | $2l$ (µm) |
|---|---|---|---|---|---|
| HEK 293 (mitochondria) | 3.7 | 1.4 | 0.052 | 0.101 | 0.104 |
| HEK 293 (lysosomes) | 3.0 | 5.8 | 0.052 | 0.137 | 0.104 |
| HeLa (mitochondria) | 7.3 | 5.8 | 0.049 | 0.123 | 0.098 |
| HeLa (beads) | 3.6 | 1.8 | 0.047 | 0.082 | 0.093 |
| HT22 (mitochondria) | 9.9 | 3.6 | 0.047 | 0.071 | 0.093 |

As a final characterisation of mitochondrial motion, we next tested whether it is ergodic. In this context, an ergodic behaviour would imply that averaging over a sufficient number of trajectories gives the same result as averaging over an individual trajectory during a sufficiently long time. It has been observed in several studies that intracellular and cell membrane phenomena present a nonergodic behaviour[37,44]. Moreover, like a non-linear mean square displacement and the decoupling of persistence exchange times, ergodicity breaking is a hallmark of glassy motion, even though it is not exclusively related to this kind of motion. We quantified the degree of ergodicity breaking based on the parameter introduced by Burov, Metzler, Barkai, and co-workers[21–23]. We did not obtain interpretable results for several of the systems we have worked with here, possibly due to a limited number of long trajectories. However, for mitochondria in HT22 cells, the results are clear and show that the motion is non-ergodic (Supplementary Figure S7). Furthermore, for beads enclosed in vesicles moving in HeLa cells, we have previously shown that their motion is also non-ergodic[3].

**CONCLUSIONS**

Here we have followed the motion of mitochondria, the energy-supplying organelles of the cell, using live-cell fluorescence microscopy. Already qualitative observation of the trajectories suggests that their motion exhibits features that are typical of those in disordered, glassy, systems. That is, the mitochondria remain within the same region for extended periods of time, stalled, before rapidly making a longer jump. Using a model that makes this physical picture more precise, we show that it provides a good description of mitochondrial motion also from a quantitative point of view. We generalise these observations to in total three different objects (mitochondria, lysosomes, and the vesicles within which nano-sized beads reside), in three different cell lines (HEK 293, HeLa, and HT22) from two different organisms (human and mouse). While the motion is quantitatively different, strikingly, in all cases we observe that it exhibits glass-like features. From model fitting we extract the characteristic lengths of the stalled motion and the jumps, as well as the waiting time before making a jump. Interestingly, the spatial extent of the stalled motion is around 0.05 µm for all systems, despite the different sizes of the organelles and despite the different cell types, while the jump length varies a bit more between systems but nevertheless remains roughly around 0.10 µm. The characteristic time before making a jump is more variable, but in all cases we observe that the time before making the first jump is different from the time before making a subsequent jump, which can be related to the concept of "ageing". We also show further evidence that the motion is non-ergodic.

It remains unknown what physical mechanisms are at the base of the glass-like behaviour. The cytosol of both prokaryotic and eukaryotic cells has been reported to behave like a soft glass[13,45,46], so this could explain why the motion in bacteria and in the cytosol of eukaryotic cells exhibits glass-like features[13,14]. However, the organelles we have followed are not just freely moving in the cytosol, but rather are mostly transported by molecular motors, like dynein and kinesin, along microtubules. It is possible that a tug-of-war mechanism that regulates the bidirectional movement of most of the cargoes along microtubules gives rise to moments in which the organelles are pulled into different directions before one of the two takes over[47,48]. Similarly, the organelle could detach from the microtubule track, thus remaining stalled, before it attaches again to make a longer move. Alternatively, crowding within the cell, due to either other organelles and/or the cytoskeleton, could also be a reason for the stalled moments, as could be a local cell energy depletion. It has also been observed that the structure of the cytoskeleton can lead to moments where the organelle remains stalled when the motor proteins meet an intersection in the mesh, and this could also form an explanation for our observations[49]. Regardless, the ubiquity of glass-like motion that we have demonstrated here suggests that it would be interesting to understand both its origins, but also its consequences.

**Author contributions**
C.Å. designed the research. A.M.D. contributed to the implementation of the research. B.C. performed all experiments and analysed all data, except as otherwise noted. O. H. performed the tracking of the lysosomes in HEK 293 cells. B.C. and C.Å. interpreted the data. B.C. and C.Å. wrote the manuscript. A.M.D. reviewed the manuscript. All authors approved the final version of the manuscript.

**Acknowledgements**
B. C. was supported by a scholarship awarded under the Advanced Materials theme of the Faculty of Science and Engineering, University of Groningen. The microscopy was performed at the University Medical Center Groningen Imaging and Microscopy Center. We thank A. Salvati (University of Groningen, Groningen, The Netherlands), for providing the HEK 293 and HeLa cells. The HT22 cells were originally provided by C. Culmsee (Philipps-Universität Marburg, Marburg, Germany) for which we are grateful. We also thank Yuequ Zhang for help with culturing and seeding the HT22 cells and Steve S. Waheeb for a preliminary analysis of lysosomal motion.

**References**

1. Philips, R., Kondev, J. & Theriot, J. *Chapter 13. A statistical view of biological dynamics. in Physical biology of the cell*. (Garland Science, 2008).
2. Alberts, B. *et al. Chapter 16. The cytoskeleton. in Molecular biology of the cell*. (Garland Science, 2008).
3. Åberg, C. & Poolman, B. Glass-like characteristics of intracellular motion in human cells. *Biophys. J.* **120**, 2355–2366 (2021).
4. Salvati, A. *et al.* Experimental and theoretical comparison of intracellular import of polymeric nanoparticles and small molecules: Toward models of uptake kinetics. *Nanomed.: Nanotechnol. Biol. Med.* **7**, 818–826 (2011).
5. Sandin, P., Fitzpatrick, L. W., Simpson, J. C. & Dawson, K. A. High-speed imaging of Rab family small GTPases reveals rare events in nanoparticle trafficking in living cells. *ACS Nano* **6**, 1513–1521 (2012).
6. Vtyurina, N., Åberg, C. & Salvati, A. Imaging of nanoparticle uptake and kinetics of intracellular trafficking in individual cells. *Nanoscale* **13**, 10436–10446 (2021).
7. Åberg, C., Varela, J. A., Fitzpatrick, L. W. & Dawson, K. A. Spatial and structural metrics for living cells inspired by statistical mechanics. *Sci. Rep.* **6**, 34457 (2016).
8. Varela, J. A., Åberg, C., Simpson, J. C. & Dawson, K. A. Trajectory-based co-localization measures for nanoparticle-cell interaction studies. *Small* **11**, 2026–2031 (2015).
9. Binder, K. & Kob, W. *Glassy materials and disordered solids: An introduction to their statistical mechanics*. (World Scientific, 2005).
10. Frey, E. & Kroy, K. Brownian motion: A paradigm of soft matter and biological physics. *Ann. Phys.* **14**, 20–50 (2005).
11. Chaudhuri, P., Berthier, L. & Kob, W. Universal nature of particle displacements close to glass and jamming transitions. *Phys. Rev. Lett.* **99**, 060604 (2007).
12. Schramma, N., Israëls, C. P. & Jalaal, M. Chloroplasts in plant cells show active glassy behavior under low light conditions. *bioRxiv* (2022). doi:10.48550/arXiv.2204.07386.
13. Parry, B. R. *et al.* The bacterial cytoplasm has glass-like properties and is fluidized by metabolic activity. *Cell* **156**, 183–194 (2014).
14. Munder, M. C. *et al.* A pH-driven transition of the cytoplasm from a fluid- to a solid-like state promotes entry into dormancy. *eLife* **5**, e09347 (2016).
15. Bernardi, P. *et al.* The functional impact of mitochondrial structure across subcellular scales. *Front. Physiol.* **11**, 1462 (2020).
16. Moore, A. S. & Holzbaur, E. L. F. Mitochondrial-cytoskeletal interactions: Dynamic associations that facilitate network function and remodeling. *Curr. Opin. Physiol.* **3**, 94–100 (2018).
17. Melkov, A. & Abdu, U. Regulation of long-distance transport of mitochondria along microtubules. *Cell. Mol. Life Sci.* **75**, 163–176 (2018).
18. Barlan, K. & Gelfand, V. I. Microtubule-based transport and the distribution, tethering, and organization of organelles. *Cold Spring Harb. Perspect. Biol.* **9**, a025817 (2017).

19. Chaudhuri, P., Gao, Y., Berthier, L., Kilfoil, M. & Kob, W. A random walk description of the heterogeneous glassy dynamics of attracting colloids. *J. Phys.: Condens. Matter* **20**, 244126 (2008).
20. Wang, B., Anthony, S. M., Chul Bae, S. & Granick, S. Anomalous yet Brownian. *Proc. Natl. Acad. Sci. U S A* **8**, 15160–15164 (2009).
21. He, Y., Burov, S., Metzler, R. & Barkai, E. Random time-scale invariant diffusion and transport coefficients. *Phys. Rev. Lett.* **101**, 058101 (2008).
22. Burov, S., Jeon, J. H., Metzler, R. & Barkai, E. Single particle tracking in systems showing anomalous diffusion: The role of weak ergodicity breaking. *Phys. Chem. Chem. Phys.* **13**, 1800–1812 (2011).
23. Metzler, R., Jeon, J.-H., Cherstvy, A. G. & Barkai, E. Anomalous diffusion models and their properties: Non-stationarity, non-ergodicity, and ageing at the centenary of single particle tracking. *Phys. Chem. Chem. Phys.* **16**, 24128–24164 (2014).
24. Jakobs, S., Stephan, T., Ilgen, P. & Brüser, C. Light microscopy of mitochondria at the nanoscale. *Annu. Rev. Biophys.* **49**, 289–308 (2020).
25. Alberts, B. *et al. Chapter 14. Energy conversion: Mitochondria and chloroplasts. in Molecular biology of the cell.* (Garland Science, 2002).
26. Valente, A. J. *et al.* Quantification of mitochondrial network characteristics in health and disease. *Adv. Exp. Med. Biol.* **1158**, 183–196 (2019).
27. Valente, A. J., Maddalena, L. A., Robb, E. L., Moradi, F. & Stuart, J. A. A simple ImageJ macro tool for analyzing mitochondrial network morphology in mammalian cell culture. *Acta Histochem.* **119**, 315–326 (2017).
28. Tinevez, J. Y. *et al.* TrackMate: An open and extensible platform for single-particle tracking. *Methods* **115**, 80–90 (2017).
29. Schindelin, J. *et al.* Fiji: An open-source platform for biological-image analysis. *Nat. Methods* **9**, 676–682 (2012).
30. Schneider, C. A., Rasband, W. S. & Eliceiri, K. W. NIH Image to ImageJ: 25 years of image analysis. *Nat. Methods* **9**, 671–675 (2012).
31. Weeks, E. R., Crocker, J. C., Levitt, A. C., Schofield, A. & Weitz, D. A. Three-dimensional direct imaging of structural relaxation near the colloidal glass transition. *Science* **287**, 627–631 (2000).
32. Marty, G. & Dauchot, O. Subdiffusion and cage effect in a sheared granular material. *Phys. Rev. Lett.* **94**, 015701 (2005).
33. Thompson, R. E., Larson, D. R. & Webb, W. W. Precise nanometer localization analysis for individual fluorescent probes. *Biophys. J.* **82**, 2775–2783 (2002).
34. Michalet, X. Mean square displacement analysis of single-particle trajectories with localization error: Brownian motion in isotropic medium. *Phys. Rev. E Stat. Nonlin. Soft Matter Phys.* **82**, 041914 (2010).
35. Binder, K. & Kob, W. *Chapter 2. Structure and dynamics of disordered matter. in Glassy materials and disordered solids: An introduction to their statistical mechanics*. (World Scientific, 2005).
36. van Hove, L. Correlations in space and time and born approximation scattering in systems of interacting particles. *Phys. Rev.* **95**, 249–262 (1954).

37. Tabei, S. M. A. *et al.* Intracellular transport of insulin granules is a subordinated random walk. *Proc. Natl. Acad. Sci. U S A* **110**, 4911–4916 (2013).
38. Witzel, P. *et al.* Heterogeneities shape passive intracellular transport. *Biophys. J.* **117**, 203–213 (2019).
39. Grady, M. E. *et al.* Intracellular nanoparticle dynamics affected by cytoskeletal integrity. *Soft Matter* **13**, 1873–1880 (2017).
40. Berthier, L. & Kob, W. The Monte Carlo dynamics of a binary Lennard-Jones glass-forming mixture. *J. Phys.: Condens. Matter* **19**, 205130 (2007).
41. Berthier, L. *et al.* Spontaneous and induced dynamic correlations in glass formers. II. Model calculations and comparison to numerical simulations. *J. Chem. Phys.* **126**, (2007).
42. Berthier, L. *et al.* Spontaneous and induced dynamic fluctuations in glass formers. I. General results and dependence on ensemble and dynamics. *J. Chem. Phys.* **126**, (2007).
43. Ballabio, A. & Bonifacino, J. S. Lysosomes as dynamic regulators of cell and organismal homeostasis. *Nat. Rev. Mol. Cell. Biol.* **21**, 101–118 (2020).
44. Weron, A. *et al.* Ergodicity breaking on the neuronal surface emerges from random switching between diffusive states. *Sci. Rep.* **7**, 5404 (2017).
45. Guo, M. *et al.* Probing the stochastic, motor-driven properties of the cytoplasm using force spectrum microscopy. *Cell* **158**, 822–832 (2014).
46. Nishizawa, K. *et al.* Universal glass-forming behavior of in vitro and living cytoplasm. *Sci. Rep.* **7**, 15143 (2017).
47. Rezaul, K. *et al.* Engineered tug-of-war between kinesin and dynein controls direction of microtubule based transport in vivo. *Traffic* **17**, 475–486 (2016).
48. Osunbayo, O. *et al.* Cargo transport at microtubule crossings: Evidence for prolonged tug-of-war between kinesin motors. *Biophys. J.* **108**, 1480–1483 (2015).
49. Scholz, M. *et al.* Cycling state that can lead to glassy dynamics in intracellular transport. *Phys. Rev. X* **6**, 011037 (2016).

# Supplementary Material to

# Extending the analogy between intracellular motion in mammalian cells and glassy dynamics

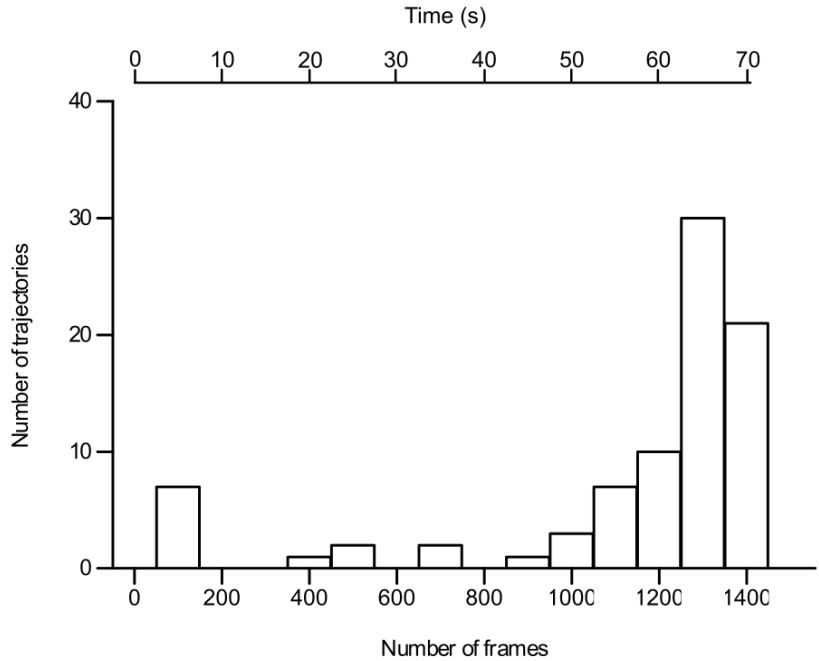

**Supplementary Figure S1: Trajectory length distribution for the mitochondria followed in HEK 293 cells.** Not all of the single mitochondria could be followed throughout all the frames (1400 frames every 50 ms for a total duration of 70 s) due to the presence of a dense mitochondrial network obscuring single mitochondria or the mitochondrion of interest moving to a different focal plane. The results here show the distribution of trajectory lengths both in number of frames (lower *x* axis) and in seconds (upper *x* axis). Total number of trajectories is 84.

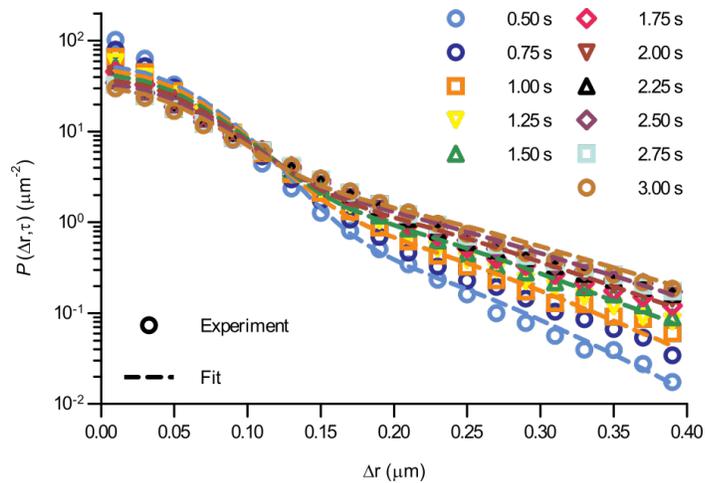

**Supplementary Figure S2: Displacement distribution of mitochondria in HEK 293 cells.** Same data as in Figure 3 of the main text, but showing more time points. (Data points) Displacement distribution of mitochondria observed experimentally for different lag times. (Dotted lines) Fits of a model describing glassy motion[11,19] to the experimental data. The four fitting parameters ($\tau_1$, $\tau_2$, $l$, $d$) were ensured to be the same for all lag times.

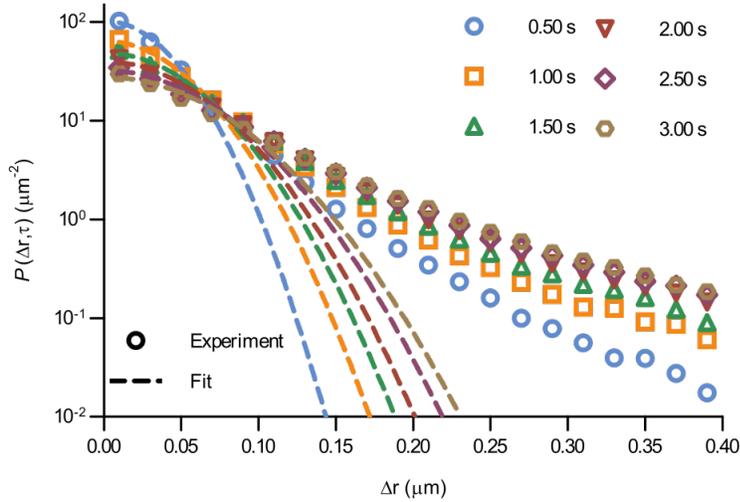

**Supplementary Figure S3: The displacement distribution of mitochondria in HEK 293 cells does not follow a Gausian displacement distribution.** (Data points) Experimental displacement distributions for a few selected times (same data as shown in Figure 3). (Dashed lines) Fit of a Gaussian distribution to the data. More specifically, we fitted a function of the form $C\exp(-\Delta r^2/A)$ to the data (note that in evaluating the experimental distributions, we have excluded a factor $2\pi\Delta r$ stemming from the two-dimensional geometry, so naturally we excluded that from the Gaussian fit as well) for $\Delta r < 0.25$ µm. For a diffusive process, the parameter $A$ would be proportional to lag time, $\tau$, but we imposed no such requirement here, rather fitting each curve separately, to give the best opportunity for good fits. However, while the short length scale behaviour ($\lesssim 0.10$ µm) is well-described by Gaussians, the tails of the distributions are clearly not. Overall, it is evident that the motion is not diffusive.

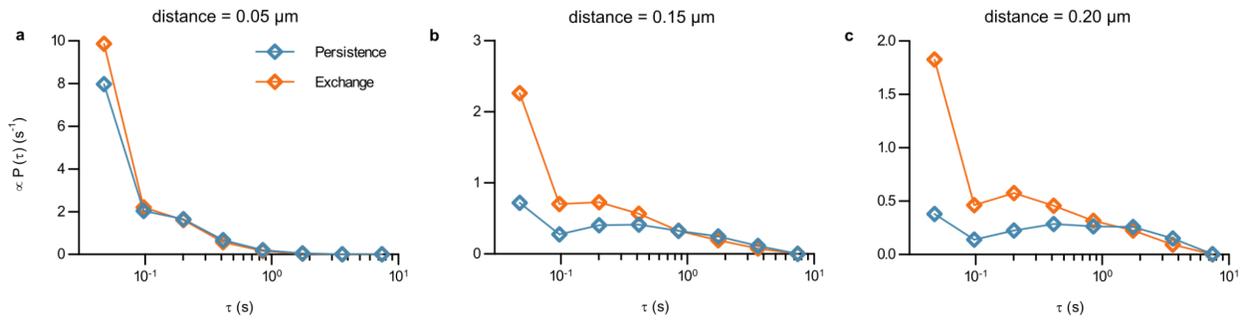

**Supplementary Figure S4: The waiting time before jumping the first time is different from the waiting time before jumping again for mitochondria in HEK 293 cells.** How long it took a mitochondrion to move a certain distance for the first time, and the time it took to move the same distance a second time having already done so once, was evaluated from the experimental data as described in Methods. To test whether the conclusion that these two times are different was independent of the choice of jump length, we performed the calculation for the lengths 0.05, 0.10, 0.15, and 0.20 µm. The choice of 0.10 µm is reported in Figure 4, while the other lengths are reported here. Overall, the results show that the waiting time before jumping the first time is different from the waiting time before jumping again, regardless of the definition of what constitutes a "jump".

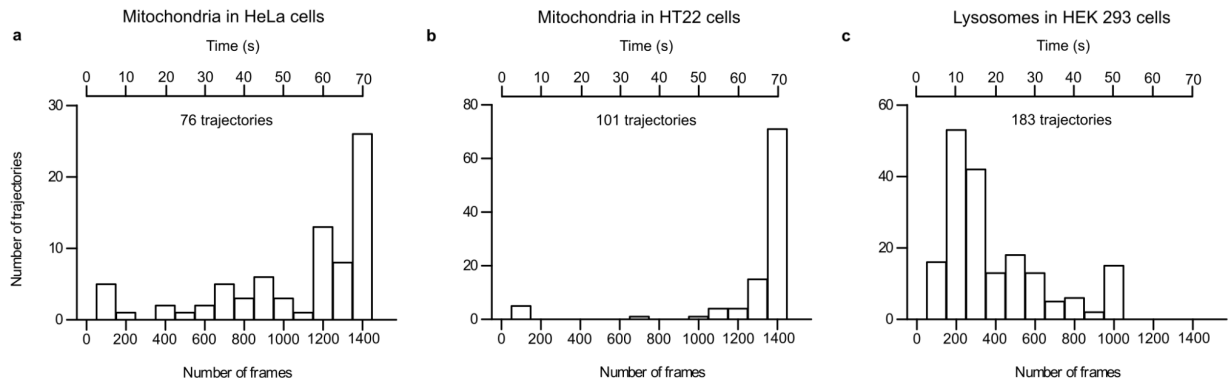

**Supplementary Figure S5: Trajectory length distribution for the organelles followed in different cell types.** Not all of the organelles could be followed throughout all the frames (1400 frames every 50 ms for a total duration of 70 s) due to the other organelles obscuring the organelle or the organelle of interest moving to a different focal plane. In the case of the lysosomes, some organelles were also lost due to photobleaching and, indeed, no lysosome was followed throughout all the frames. The results here show the distribution of trajectory lengths both in number of frames (lower *x* axis) and in seconds (upper *x* axis). Total number of trajectories is as reported in the relevant panel.

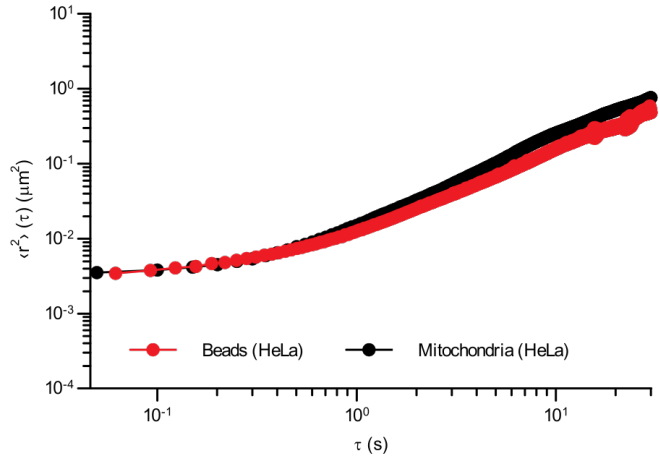

**Supplementary Figure S6: Comparison of the mean square displacement of beads within vesicles and mitochondria in the same cell type (HeLa cells).** The mean square displacement was calculated from the trajectories, averaging both over time and trajectories (ensemble). Error bars represent standard error of the mean, but are too small to be visible. (Black) Beads within vesicles, reproduced from previous literature[3]. (Red) Mitochondria, from this work. Same data as presented in Figure 5 (upper row). *N. B.* the log-log scale.

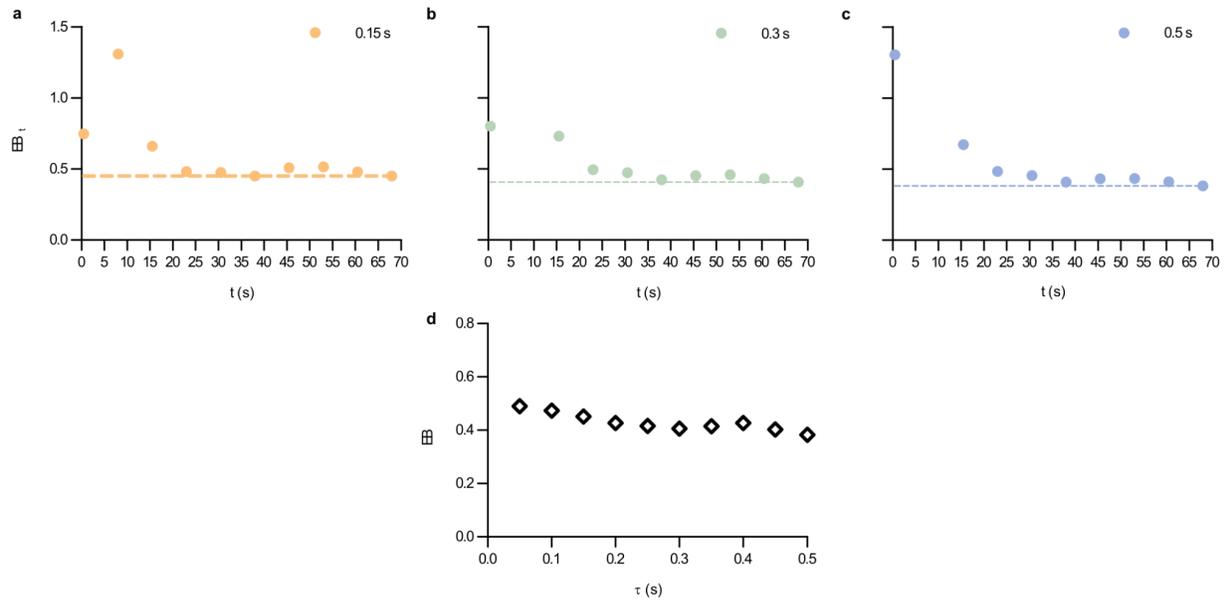

**Supplementary Figure S7: The motion of mitochondria in HT22 cells is non-ergodic.** We characterised whether their motion was ergodic or not using the ergodicity breaking parameter, EB, introduced in previous literature[21–23]. This parameter in general depends on lag time, τ, but for ergodic motion it is independent of lag time and equal to 0. Furthermore, this parameter is defined as an infinite time limit, so we started by evaluating a time-dependent version of the parameter, $EB_t$, that is, we quantified the parameter which in the infinite time limit becomes the ergodicity breaking parameter, EB. (a–c) (Datapoints) Parameter, $EB_t$, that tends to the ergodicity breaking parameter in the infinite time limit for three exemplar lag times, τ = 0.15, 0.30 and 0.50 s. (Dashed lines) Value after 70 s to visualise the convergence of $EB_t$. The fact that the values remain roughly the same after 20–30 s suggests that using $EB_t$ for *t* = 70 s is a reasonable approximation of the ergodicity breaking parameter EB, which, in principle, is only defined as the infinite time limit. (d) Ergodicity breaking parameter as function of the lag time, τ. Only lag times where we observed a convergence (as in panels a–c) where included. For all lag times, we observe an ergodicity breaking parameter close to 0.4. This is clearly distinct from 0, showing that the motion is non-ergodic.

24